\documentclass[prl,twocolumn,showpacs,amsmath,floatfix]{revtex4}
\usepackage{graphicx}
\usepackage{bm}

\begin{document}
\title{Measurement of the Casimir force between a gold sphere and silicon surface with nanoscale trench arrays}

\author{H. B. Chan}
\email{hochan@phys.ufl.edu}
\author{Y. Bao}
\author{J. Zou}

\affiliation{Department of Physics, University of Florida, Gainesville, Florida 32611}

\author{R. A. Cirelli}
\author{F. Klemens}
\author{W. M. Mansfield}
\author {C. S. Pai}

\affiliation{Bell Laboratories, Lucent Technologies, Murray Hill, NJ 07974}

%
\begin{abstract}
We report measurements of the Casimir force between a gold sphere and a silicon surface with an array of nanoscale, rectangular corrugations using a micromechanical torsional oscillator. At distance between 150$\:$nm and 500$\:$nm, the measured force shows significant deviations from the pairwise additive formulism, demonstrating the strong dependence of the Casimir force on the shape of the interacting bodies. The observed deviation, however, is smaller than the calculated values for perfectly conducting surfaces, possibly due to the interplay between finite conductivity and geometry effects.
\end{abstract}
\pacs{03.70.+k, 12.20.Fv, 12.20.Ds, 42.50.Lc}

\maketitle

The Casimir force is the interaction between neutral conductors that can be understood as resulting from the alteration of the zero point energy of the electromagnetic field in the presence of boundaries \cite{Casimir1948}. For two perfect metallic planar surfaces, the force is attractive and is given by $F_c=\pi ^2\hbar cA/240z^4$, where $c$ is the speed of light, $\hbar$ is the Planck's constant$/2\pi$, $z$ is the separation between the plates and $A$ is the area of the plates. There exists a close connection between the Casimir force between conductors and the van der Waals (vdW) force between molecules. For the former, the quantum fluctuations are often associated with the vacuum electromagnetic field, while the latter commonly refers to the interaction between fluctuating dipoles. In simple geometries such as two parallel planes, the Casimir force can be interpreted as an extension of the vdW force in the retarded limit. The interaction between molecules in the two plates is summed to yield the total force. However, such summation of the vdW force is not always valid for extended bodies because the vdW force is not pairwise additive. The interaction between two molecules is affected by the presence of a third molecule. One important characteristic of the Casimir force is its strong dependence on geometry \cite{Rodriguez2007}. The Casimir energy for a conducting spherical shell \cite{Boyer1968} or a rectangular box \cite{Maclay2000,Gusso2006} has been calculated to have opposite sign to parallel plates. Whether such geometries exhibit repulsive Casimir forces remains a topic of current interest \cite{Hertzberg2005}. 

In recent years, there have been a number of precision measurements of the Casimir force \cite{Lamoreaux1997,Mohideen1998,Ederth2000,Chan2001,Bressi2002,Decca2003,Chen2006,Decca2005,Munday2007}. These experiments yield agreement with the theoretical calculations to accuracies of better than one percent when nonideal behavior of the metallic surfaces \cite{Lifshitz1956, Klimchitskaya1999,Neto2005} are taken into account. The vast majority of force measurements were performed between a sphere and a flat plate, two flat plates or two cylinders.  For these simple geometries, the Casimir force is not expected to show significant deviations from the pairwise additive approximation (PAA) at small separations. There has only been one experiment that involved surfaces of other geometries, where the Casimir force is measured between a sphere and a plate with small sinusoidal corrugations \cite{Roy1999}. While this measurement shows deviations from PAA, the interpretation of the deviation is still controversial. For example, the Casimir force for sinusoidal corrugations on perfect conductors was calculated without the assumption of PAA \cite{Emig2001}. The deviation from PAA is found to be strong only when the ratio of the separation to the periodicity of corrugation $\lambda$ is large. In ref. \cite{Roy1999}, $\lambda$ is not small enough for deviations from PAA to be significant. It is suggested that lateral movement of the two surfaces may be able to account for the deviations \cite{Klimchitskaya2000}.

\begin{figure}%
\includegraphics[angle=0]{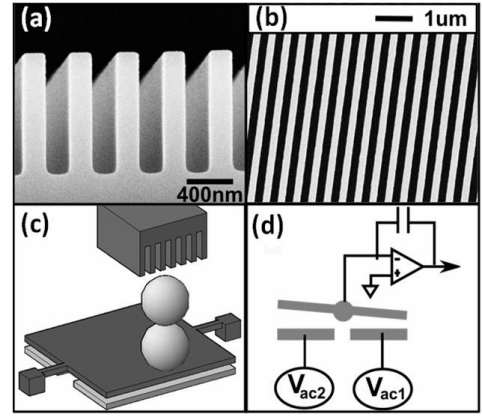} 
\caption{\label{fig:1} (a) Cross section of rectangular trenches in silicon, with periodicity of 400$\:$nm and depth of 0.98$\:$$\mu$m (sample B). (b) Top view of the structure. (c) Schematic of the experimental setup (not to scale) including the micromechanical torsional oscillator, gold spheres and silicon trench array. (d) Measurement scheme with electrical connections. Excitation voltages $V_{ac1}$ and $V_{ac2}$ are applied to the bottom electrodes.}
\end{figure}

In this letter, we report measurements of the Casimir force between nanostructured silicon surfaces and a gold sphere. One of the interacting objects consists of a silicon surface with nanoscale, high aspect ratio rectangular corrugations. The other surface is a gold-coated glass sphere attached onto a micromechanical torsional oscillator. Lateral movements of the surfaces are avoided by positioning the corrugations perpendicular to the torsional axis. The Casimir force gradient is measured from the shifts in the resonant frequency of the oscillator at distance between 150$\:$nm and 500$\:$nm. Deviations of up to $20\%$ from PAA is observed, demonstrating the strong geometry dependence of the Casimir force. The measured deviation is, however, about a factor of 2 smaller than deviations expected for perfectly conducting surfaces \cite{Buscher2004}.

Figure$\:$\ref{fig:1}a shows a cross section of an array of rectangular corrugations with period of 400$\:$nm (sample B) fabricated on a highly p-doped silicon substrate. Two other samples, one with period 1$\:$$\mu$m (sample A) and the other with a flat surface, are also fabricated. The fabrication procedure started with a layer of silicon oxide (0.2$\:$$\mu$m) deposited onto a blank silicon wafer by chemical vapor deposition. Lithography was performed with a deep ultra-violet stepper followed by reactive ion etching to transfer the pattern from the photoresist to the silicon oxide. Trenches with depths $t=2a$ ($\sim$1$\:$$\mu$m) were then created using deep reactive ion etching using the oxide as etch mask. A continuous etch and deposit recipe was used to yield smooth side walls at $90.3\,^{\circ}$ and $91.0\,^{\circ}$ to the top surface respectively for samples A and B. Residual hydrocarbons were removed using an oxygen plasma etch. Finally, the oxide etch mask is removed using hydrofluoric acid. In order to ensure that the optical properties of the silicon are identical, all samples were fabricated on the same wafer and later diced into 0.7mm by 0.7mm pieces for the force measurement. 

The geometry of nanoscale, rectangular trenches was chosen because the Casimir force on such structures are expected to exhibit large deviations from PAA. We consider the interaction between the trench array and a parallel flat surface at distance $z$ from the top surface of the trenches. In the pairwise additive picture, this interaction is a sum of two contributions: the volume from the top surface to the bottom of the trench and the volume below the bottom of the trench. The latter component is negligible because the distance to the other surface is more than 1$\:$$\mu$m, larger than the distance range at which Casimir forces can be detected in our experiment. For a trench array of $50\%$ duty cycle, the former component yields exactly half of the interaction between two flat surfaces $F_{flat}$ regardless of the periodicity because half of the material is removed \cite{Buscher2004}. In practice, the trench arrays are created with duty cycle close to but not exactly at $50\%$. Under PAA, the total force is equal to $pF_{flat}$, where $p$ is the fraction of solid volume. The calculation of the Casimir force in such corrugated surfaces, in contrast, is highly non-trivial. While perturbative treatments \cite{Emig2001} are valid for smooth profiles with small local curvature, they are impractical for the deep, rectangular corrugations. Using a different approach based on path integrals, B\"{u}scher and Emig \cite{Buscher2004} calculated the Casimir force for the corrugated geometry made of perfect conductors. Strong deviations from PAA were obtained when the ratio $z/\lambda$ is large. In the limit when $\lambda$ goes to zero, the force on a trench array approaches the value between flat surfaces, leading to deviations from PAA by a factor of 2. Such large deviations occur because the Casimir force is associated with confined electromagnetic modes with wavelength comparable to the separation between the interacting objects. When $\lambda<<z$, these modes fail to penetrate into the trenches, rendering the Casimir force on the corrugated surface equal to a flat one.  

We measure the gradient of the Casimir force on the silicon trench arrays using a gold-coated sphere attached to a micromechanical torsional oscillator \cite{Chan2001}. The oscillator consists of a 3.5$\:$$\mu$m thick, 500$\:$$\mu$m square silicon plate suspended by two torsional rods. As shown in Fig.$\:$\ref{fig:1}c, two glass spheres, each with radius $R$ of 50$\:$$\mu$m, are stacked and attached by conductive epoxy onto the oscillator \cite{Decca2005} at a distance of  $b =  210\:\mu$m from the rotation axis. The large distance ($\sim 200\:\mu$m) between the oscillator plate and the corrugated surface ensures that the attraction between them is negligible and only the interaction between the top sphere and the corrugated surface is measured. Before attachment, a layer of gold with thickness 4000$\:$A is sputtered onto the spheres. Two electrodes are located between the plate and the substrate. Torsional oscillations in the plate are excited when the voltage on one of the electrodes is modulated at the resonant frequency of the oscillator ($f_0$ = 1783 Hz, quality factor $Q = 32000$). For detecting the oscillations, additional ac voltages at amplitude of 100$\:$mV and frequency of 102$\:$kHz is applied to measure the capacitance change between the top plate and the electrodes. A phase locked loop is used to track the shifts in the resonance frequency \cite{Chan2001} as the sphere approaches the other silicon plate through extension of a closed-loop piezoelectric actuator. As shown in Fig.$\:$\ref{fig:1}c, the movable plate is positioned so that its torsional axis is perpendicular to the trench arrays in the other silicon surface. Such an arrangement eliminates motion of the movable plate in response to lateral Casimir forces \cite{Klimchitskaya2000,Chen2002} because the spring constant for translation along the torsional axis is orders of magnitude larger than the orthogonal direction in the plane of the substrate.

To prepare the silicon surface \cite{Chen2006} for force measurement, hydrofluoric acid is used to remove the native oxide on the surface of the silicon chip. The hydrofluoric acid also passivates the silicon surface to temporarily prevent oxide formation at ambient pressure. In the next step, the silicon chip is baked at $120\,^\circ$C for 15 minutes to eliminate residual water that might have accumulated in the trenches. The silicon chip is then positioned to within a few micrometers from the gold sphere and the chamber is immediately evacuated to a pressure of $10^{-6}$ torr by a dry roughing pump and a turbo pump.

\begin{figure}%
\includegraphics[width=2.6in]{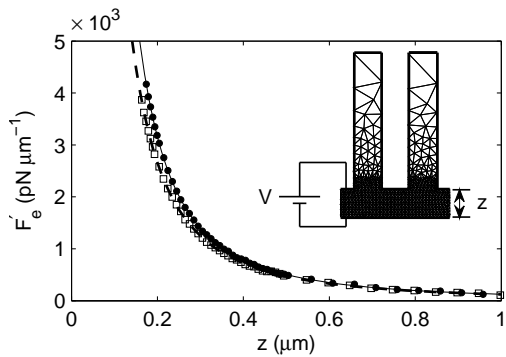} 
\caption{\label{fig:2} Gradient of the electrostatic force ($V = V_0 + 300\:$mV) on the flat silicon surface (solid circles) and corrugated silicon structure (hollow squares). The solid line is a fit using Eq. \ref{eq:2} for the flat surface. The dashed line is a fit using the force gradient from finite element analysis for the corrugated structure. Inset: The space between the corrugated structure and a flat surface is divided into triangular mesh to solve the Poisson equation in 2D ($z = 200\:$nm).} 
\end{figure}

For small oscillations where nonlinear effects can be neglected, the resonant frequency shift $\Delta f$ of the oscillator is proportional to the gradient of the force $F^{\prime}(z)$ between the surfaces: 
\begin{equation}
\label{eq:1}
\Delta f=CF^{\prime}(z),
\end{equation}
where $C=-b^2/8\pi ^2f_0I$ and $I$ is the moment of inertia of the oscillator. The distance $z$ is given by $z_0-z_{piezo}-b\theta$, where $z_0$ is the initial separation between the two surfaces, $z_{piezo}$ is the extension of the piezoelectric actuator and $b \theta$ is the modification of the separation due to rotation of the top plate to angle $\theta$. To calibrate $z_0$ and the proportionality constant $C$, a dc voltage $V$ is applied to the silicon plate with the gold sphere electrically grounded. The electrostatic force between a sphere and a flat plate is given by:
\begin{equation}
\label{eq:2}
F_e = 2\pi \epsilon_0(V-V_0)\sum _{n=1}^{\infty}\frac{[\coth (\alpha)-n\coth (n\alpha)]}{\sinh(n\alpha)},
\end{equation}
where $\cosh \alpha = 1 + d/R$ and $\epsilon _0$ is the permittivity of vacuum. The residual voltage $V_0$ is measured to be $\sim -0.43\:V$ by determining $V$ at which $\Delta f$ attains minimum at fixed $z$. For 100$\:$nm $< z <$ 2$\:\mu$m, $V_0$ is found to vary by less than 3 mV. In Fig.$\:$\ref{fig:2}, the solid line represents a fit to the measured electrostatic force gradient for a flat surface. The contribution of the Casimir force gradient to the measured frequency shift ($< 4\%$ at the smallest $z$) has been subtracted. By averaging the fitted values for six sets of data with $V$ ranging from $V_0 + 245$$\:$mV to $V_0 + 300$$\:$mV, $C$ is determined to be 628$\:\pm\:$5$\:$m$\:N^{-1}\:s^{-1}$. The voltages are chosen to be larger than $V_0$ to avoid depleting the surface of the p-doped silicon with charge carriers. A similar calibration procedure is performed on the silicon trench arrays (dashed line in Fig.$\:$\ref{fig:2}). Since there is no analytic expression for the trench geometry, the electrostatic force is calculated by solving Poisson's equation in 2D using finite element analysis. As shown in the inset to Fig.$\:$\ref{fig:2}, the boundary conditions are set by maintaining a fixed potential between the trench array and a flat surface, with the volume between them divided into $N>10,000$ triangles. Since $R >> z$, the proximity force approximation $F_{sc} = 2 \pi R E_{fc}$ can be used to obtain the force $F_{sc}$ between a sphere and a corrugated surface from the electrostatic energy $E_{fc}$ between a flat surface and a corrugated surface. To ensure that $N$ is sufficiently large, we checked that the calculated force varies by less than $0.1\%$ even when $N$ is doubled. 

The dimensions of the trench array used in the calculation were obtained from the scanning electron micrographs: the fraction of solid volume $p$ is determined from the percentage of bright pixels in the top view similar to Fig.$\:$\ref{fig:1}b. Ten pictures at different locations with area 30$\:\mu$m by 50$\:\mu$m for sample A and 20$\:\mu$m by 35$\:\mu$m  for sample B were used to calculate $p_A$ and $p_B$ to be $0.478\pm0.002$ and $0.510\pm0.001$ respectively. The cross sectional view yields the trench depth ($t_A =0.98\:\mu$m and $t_B = 1.07\:\mu$m). As shown in Fig. 1a, while we have optimized the fabrication process so that the top corner of the trenches has a sharp rectangular shape, the bottom sections show certain degree of rounding. In all analysis described here, the trenches are assumed to have perfect rectangular shape. The validity of such approximation is justified by the insensitivity of the calculated electrostatic force on the depth of the trenches. Varying the depth of the trenches by 10\% produces less than 0.01\% change in the calculated force.

\begin{figure}%
\includegraphics[angle=0]{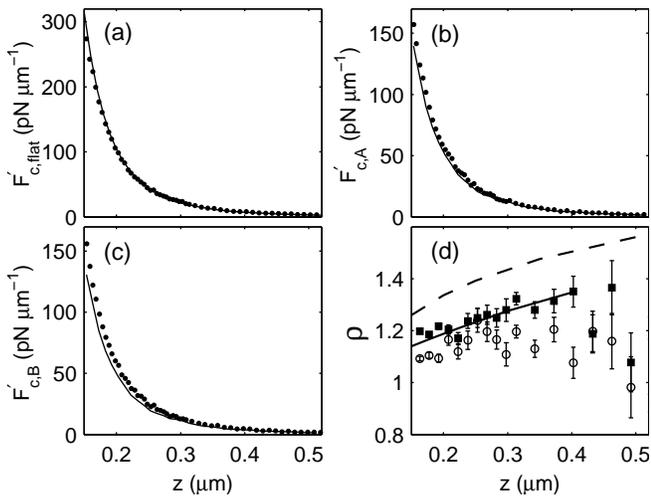} 
\caption{\label{fig:3} Measured Casimir force gradient between the same gold sphere and (a) a flat silicon surface, $F^{\prime}_{c,flat}$, (b) sample A, $F^{\prime}_{c,A}$ ($\lambda=1\:\mu$m) and (c) sample B, $F^{\prime}_{c,B}$ ($\lambda=400\:$nm). In (a), the line represents the theoretical Casimir force gradient including finite conductivity and surface roughness corrections. In (b) and (c), the lines represent the force gradients expected from PAA ($pF^{\prime}_{c,flat}$). (d) Ratio $\rho$ of the measured Casimir force gradient to the force gradient expected from PAA, for samples A ($\lambda/a=1.87$, hollow circles) and B ($\lambda/a=0.82$, solid squares) respectively. Theoretical values \cite{Buscher2004} for perfectly conducting surfaces are plotted as the solid ($\lambda/a=2$) and dashed lines ($\lambda/a=1$).} 
\end{figure}

By setting $V$ equal to $V_0$, the Casimir force gradient $F^{\prime}_{c,flat}$ is measured between the gold sphere and a flat silicon surface (solid circles in Fig.$\:$\ref{fig:3}a) obtained from same wafer on which the corrugated samples A and B were fabricated. The main source of uncertainty in the measurement ($\sim 0.64\:$pN$\mu$m$^{-1}$ at $z = 300\:$nm) originates from the thermomechanical fluctuations of the micromechanical oscillator. As the distance decreases, the oscillation amplitude is reduced to prevent the oscillator from entering the nonlinear regime \cite{Chan2001}. At distances below 150$\:$ nm, the oscillation amplitude becomes too small for reliable operation of the phase-locked loop. In Fig.$\:$\ref{fig:3}a, the line represents the theoretical force gradient between the gold sphere and the flat silicon surface, including both the finite conductivity and roughness corrections. Lifshiftz's expressions \cite{Lifshitz1956,Klimchitskaya1999,Chen2006} are used to take into account the finite conductivity. For the gold surface, tabulated values of the optical properties \cite{Palik1985} were used. For the silicon surface, the tabulated values were further modified by the concentration of carriers ($2 \times 10^{18}\:$cm$^{-3}$) determined from the dc conductivity of the wafer ($0.028\: \Omega \:$cm) \cite{Chen2006}. Using an atomic force microscope (NTMDT), the main contribution to the roughness is found to originate from the gold surface ($\sim 4\:$nm rms) rather than the silicon wafer ($\sim 0.6\:$nm rms). The roughness correction \cite{Klimchitskaya1999,Neto2005} is taken into account by the geometrical averaging method \cite{Chen2006}.
  
The Casimir force gradients  $F^{\prime}_{c,A}$ and $F^{\prime}_{c,B}$ between the same gold sphere and the corrugated samples A and B were then measured and plotted in Figs. \ref{fig:3}b and \ref{fig:3}c. As described earlier, under PAA, the forces on the trench arrays (where $z$ is measured from the top of the corrugated surface)  are equal to the force on a flat surface multiplied by the fractional volumes $p_A$ and $p_B$. The solid lines in Figs. \ref{fig:3}b and \ref{fig:3}c represent the corresponding force gradients, $p_A\,F^{\prime}_{c,flat}$ and $p_B\,F^{\prime}_{c,flat}$ respectively. Measurement of the force gradient was repeated three times for each sample, yielding results that are consistent within the measurement uncertainty. To analyze the deviations from PAA, the ratios $\rho _A=F^{\prime}_{c,A}/p_A\,F^{\prime}_{c,flat}$ and $\rho _B=F^{\prime}_{c,B}/p_B\,F^{\prime}_{c,flat}$ are plotted in Fig. \ref{fig:3}d. The ratio $\rho$ equals one if PAA is valid. For sample A with $\lambda/a=1.87$, where $a$ is half the depth of the trenches, the measured force deviates from PAA by $\sim 10\%$. In sample B with $\lambda/a=0.82$, the deviation increases to $\sim 20\%$. For 150$\:$nm $< z <$ 250$\:$nm, the measured Casimir force gradient in both samples show clear deviations from PAA. At larger distances, the uncertainty increases considerably as the force gradient decreases. 

We compare our experimental results on silicon structures to calculations by B\"{u}scher and Emig \cite{Buscher2004} on perfect conductors (solid and dashed lines in Fig. \ref{fig:3}d). In this calculation, the Casimir force between a flat surface and a corrugated structure with $p = 0.5$ was determined for a range of $\lambda /a$ using a path integral approach. Since $R >> z$, the proximity force approximation allows a direct comparison of our measured force gradient using a sphere and the predicted force that involved a flat surface. The measured deviation in sample B is larger than sample A, in agreement with the notion that geometry effects become stronger as $\lambda /a$ decreases. However, the measured deviations from PAA are smaller than the predicted values by about $50\%$, significantly exceeding the measurement uncertainty for 150$\:$nm $< z <$ 250$\:$nm. Such discrepancy is, to a certain extent, expected as a result of the interplay between finite conductivity and geometry effects. The relatively large value of the skin depth in silicon ($\sim 11\:$nm at wavelength of 300$\:$nm) could reduce the deviations from PAA. So far, exact computation of the Casimir force including both finite conductivity and geometry effects for strongly deformed rectangular trenches has not yet been performed. It is a nontrivial problem that warrants further theoretical analysis.

In conclusion, we measured the Casimir force gradient between a spherical gold surface and a silicon surface with an array of deep, rectangular trenches. Deviation from the PAA is clearly observed and found to increase as $\lambda /a$ decreases, consistent with the strong boundary dependence of the Casimir force. However, the measured deviation is smaller than the calculated value on perfect conductors. 

This work was supported by DOE DE-FG02-05ER46247 and NSF DMR-0645448. We thank A. Hanke for useful discussions and T. Emig for supplying the calculated values of the Casimir force on corrugations in perfect metal.







\begin{thebibliography}{23}
\expandafter\ifx\csname natexlab\endcsname\relax\def\natexlab#1{#1}\fi
\expandafter\ifx\csname bibnamefont\endcsname\relax
  \def\bibnamefont#1{#1}\fi
\expandafter\ifx\csname bibfnamefont\endcsname\relax
  \def\bibfnamefont#1{#1}\fi
\expandafter\ifx\csname citenamefont\endcsname\relax
  \def\citenamefont#1{#1}\fi
\expandafter\ifx\csname url\endcsname\relax
  \def\url#1{\texttt{#1}}\fi
\expandafter\ifx\csname urlprefix\endcsname\relax\def\urlprefix{URL }\fi
\providecommand{\bibinfo}[2]{#2}
\providecommand{\eprint}[2][]{\url{#2}}

\bibitem[{\citenamefont{Casimir}(1948)}]{Casimir1948}
\bibinfo{author}{\bibfnamefont{H.~B.~G.} \bibnamefont{Casimir}},
  \bibinfo{journal}{Proc. Kon. Ned. Akad. Wet.} \textbf{\bibinfo{volume}{51}},
  \bibinfo{pages}{793} (\bibinfo{year}{1948}).

\bibitem[{\citenamefont{Rodriguez et~al.}(2007)\citenamefont{Rodriguez,
  Ibanescu, Iannuzzi, Capasso, Joannopoulos, and Johnson}}]{Rodriguez2007}
\bibinfo{author}{\bibfnamefont{A.}~\bibnamefont{Rodriguez}} {\it et al.},
  \bibinfo{journal}{Phys. Rev. Lett.}
  \textbf{\bibinfo{volume}{99}}, \bibinfo{pages}{080401}
  (\bibinfo{year}{2007}).

\bibitem[{\citenamefont{Boyer}(1968)}]{Boyer1968}
\bibinfo{author}{\bibfnamefont{T.~H.} \bibnamefont{Boyer}},
  \bibinfo{journal}{Phys. Rev.} \textbf{\bibinfo{volume}{174}},
  \bibinfo{pages}{1764} (\bibinfo{year}{1968}).

\bibitem[{\citenamefont{Maclay}(2000)}]{Maclay2000}
\bibinfo{author}{\bibfnamefont{G.~J.} \bibnamefont{Maclay}},
  \bibinfo{journal}{Phys. Rev. A} \textbf{\bibinfo{volume}{61}},
  \bibinfo{pages}{052110} (\bibinfo{year}{2000}).
  
\bibitem[{\citenamefont{Gusso and Schmidt}(2006)}]{Gusso2006}
\bibinfo{author}{\bibfnamefont{A.}~\bibnamefont{Gusso}} \bibnamefont{and}
  \bibinfo{author}{\bibfnamefont{A.~G.~M.} \bibnamefont{Schmidt}},
  \bibinfo{journal}{Braz. J. Phys.} \textbf{\bibinfo{volume}{36}},
  \bibinfo{pages}{168} (\bibinfo{year}{2006}).
  
\bibitem[{\citenamefont{Hertzberg et~al.}(2005)}]{Hertzberg2005}
\bibinfo{author}{\bibfnamefont{M.~P.} \bibnamefont{Hertzberg}} {\it et al.},
  \bibinfo{journal}{Phys. Rev. Lett.} \textbf{\bibinfo{volume}{95}},
  \bibinfo{pages}{250402} (\bibinfo{year}{2005}).

\bibitem[{\citenamefont{Lamoreaux}(1997)}]{Lamoreaux1997}
\bibinfo{author}{\bibfnamefont{S.~K.} \bibnamefont{Lamoreaux}},
  \bibinfo{journal}{Phys. Rev. Lett.} \textbf{\bibinfo{volume}{78}},
  \bibinfo{pages}{5} (\bibinfo{year}{1997}).

\bibitem[{\citenamefont{Mohideen and Roy}(1998)}]{Mohideen1998}
\bibinfo{author}{\bibfnamefont{U.}~\bibnamefont{Mohideen}} \bibnamefont{and}
  \bibinfo{author}{\bibfnamefont{A.}~\bibnamefont{Roy}},
  \bibinfo{journal}{Phys. Rev. Lett.} \textbf{\bibinfo{volume}{81}},
  \bibinfo{pages}{4549} (\bibinfo{year}{1998}).

\bibitem[{\citenamefont{Ederth}(2000)}]{Ederth2000}
\bibinfo{author}{\bibfnamefont{T.}~\bibnamefont{Ederth}},
  \bibinfo{journal}{Phys. Rev. A} \textbf{\bibinfo{volume}{62}},
  \bibinfo{pages}{062104} (\bibinfo{year}{2000}).

\bibitem[{\citenamefont{Chan et~al.}(2001{\natexlab{a}})\citenamefont{Chan,
  Aksyuk, Kleiman, Bishop, and Capasso}}]{Chan2001}
\bibinfo{author}{\bibfnamefont{H.~B.} \bibnamefont{Chan}} {\it et al.},
  \bibinfo{journal}{Science} \textbf{\bibinfo{volume}{291}},
  \bibinfo{pages}{1941} (\bibinfo{year}{2001}{\natexlab{a}}),
  \bibinfo{journal}{Phys. Rev. Lett.} \textbf{\bibinfo{volume}{87}},
  \bibinfo{pages}{211801} (\bibinfo{year}{2001}{\natexlab{b}}).

\bibitem[{\citenamefont{Bressi et~al.}(2002)\citenamefont{Bressi, Carugno,
  Onofrio, and Ruoso}}]{Bressi2002}
\bibinfo{author}{\bibfnamefont{G.}~\bibnamefont{Bressi}} {\it et al.},
  \bibinfo{journal}{Phys. Rev. Lett.} \textbf{\bibinfo{volume}{88}},
  \bibinfo{pages}{041804} (\bibinfo{year}{2002}).

\bibitem[{\citenamefont{Decca et~al.}(2003)\citenamefont{Decca, Lopez,
  Fischbach, and Krause}}]{Decca2003}
\bibinfo{author}{\bibfnamefont{R.~S.} \bibnamefont{Decca}} {\it et al.},
  \bibinfo{journal}{Phys. Rev. Lett.}
  \textbf{\bibinfo{volume}{91}}, \bibinfo{pages}{050402}
  (\bibinfo{year}{2003}).

\bibitem[{\citenamefont{Decca et~al.}(2005)}]{Decca2005}
\bibinfo{author}{\bibfnamefont{R.~S.} \bibnamefont{Decca}} {\it et al.},
  \bibinfo{journal}{Phys. Rev. Lett.}
  \textbf{\bibinfo{volume}{94}}, \bibinfo{pages}{240401}
  (\bibinfo{year}{2005}).
  
\bibitem[{\citenamefont{Chen et~al.}(2006)\citenamefont{Chen, Klimchitskaya,
  Mostepanenko, and Mohideen}}]{Chen2006}
\bibinfo{author}{\bibfnamefont{F.}~\bibnamefont{Chen}},
  \bibinfo{author}{\bibfnamefont{G.~L.} \bibnamefont{Klimchitskaya}},
  \bibinfo{author}{\bibfnamefont{V.~M.} \bibnamefont{Mostepanenko}},
  \bibnamefont{and} \bibinfo{author}{\bibfnamefont{U.}~\bibnamefont{Mohideen}},
  \bibinfo{journal}{Phys. Rev. Lett.} \textbf{\bibinfo{volume}{97}},
  \bibinfo{pages}{170402} (\bibinfo{year}{2006}).

\bibitem[{\citenamefont{Munday and Capasso}(2007)}]{Munday2007}
\bibinfo{author}{\bibfnamefont{J.~N.} \bibnamefont{Munday}} \bibnamefont{and}
  \bibinfo{author}{\bibfnamefont{F.}~\bibnamefont{Capasso}},
  \bibinfo{journal}{Phys. Rev. A} \textbf{\bibinfo{volume}{75}},
  \bibinfo{pages}{060102(R)} (\bibinfo{year}{2007}).

\bibitem[{\citenamefont{Lifshitz}(1956)}]{Lifshitz1956}
\bibinfo{author}{\bibfnamefont{E.~M.} \bibnamefont{Lifshitz}},
  \bibinfo{journal}{Sov. Phys. JETP} \textbf{\bibinfo{volume}{2}},
  \bibinfo{pages}{73} (\bibinfo{year}{1956}).

\bibitem[{\citenamefont{Klimchitskaya et~al.}(1999)\citenamefont{Klimchitskaya,
  A.~Roy, and Mostepanenko}}]{Klimchitskaya1999}
\bibinfo{author}{\bibfnamefont{G.~L.} \bibnamefont{Klimchitskaya}},
  \bibinfo{author}{\bibfnamefont{A.} ~\bibnamefont{Roy}},
  \bibinfo{author}{\bibfnamefont{U.}~\bibnamefont{Mohideen}},
  \bibnamefont{and} \bibinfo{author}{\bibfnamefont{V.~M.} \bibnamefont{Mostepanenko}},
  \bibinfo{journal}{Phys. Rev. A}
  \textbf{\bibinfo{volume}{60}}, \bibinfo{pages}{3487} (\bibinfo{year}{1999}).

\bibitem[{\citenamefont{Neto et~al.}(2005)\citenamefont{Neto, Lambrecht, and
  Reynaud}}]{Neto2005}
\bibinfo{author}{\bibfnamefont{P.~A.~M.} \bibnamefont{Neto}},
  \bibinfo{author}{\bibfnamefont{A.}~\bibnamefont{Lambrecht}},
  \bibnamefont{and} \bibinfo{author}{\bibfnamefont{S.}~\bibnamefont{Reynaud}},
  \bibinfo{journal}{Phys. Rev. A} \textbf{\bibinfo{volume}{72}},
  \bibinfo{pages}{012115} (\bibinfo{year}{2005}).

\bibitem[{\citenamefont{Roy and Mohideen}(1999)}]{Roy1999}
\bibinfo{author}{\bibfnamefont{A.}~\bibnamefont{Roy}} \bibnamefont{and}
  \bibinfo{author}{\bibfnamefont{U.}~\bibnamefont{Mohideen}},
  \bibinfo{journal}{Phys. Rev. Lett.} \textbf{\bibinfo{volume}{82}},
  \bibinfo{pages}{4380} (\bibinfo{year}{1999}).

\bibitem[{\citenamefont{Emig et~al.}(2001)\citenamefont{Emig, Hanke,
  Golestanian, and Kardar}}]{Emig2001}
\bibinfo{author}{\bibfnamefont{T.}~\bibnamefont{Emig}} {\it et al.},
  \bibinfo{journal}{Phys. Rev. Lett.} \textbf{\bibinfo{volume}{87}},
  \bibinfo{pages}{260402} (\bibinfo{year}{2001}).

\bibitem[{\citenamefont{Klimchitskaya et~al.}(2000)\citenamefont{Klimchitskaya,
  Zanette, and Caride}}]{Klimchitskaya2000}
\bibinfo{author}{\bibfnamefont{G.~L.} \bibnamefont{Klimchitskaya}},
  \bibinfo{author}{\bibfnamefont{S.~I.} \bibnamefont{Zanette}},
  \bibnamefont{and} \bibinfo{author}{\bibfnamefont{A.~O.}
  \bibnamefont{Caride}}, \bibinfo{journal}{Phys. Rev. A}
  \textbf{\bibinfo{volume}{63}}, \bibinfo{pages}{014101}
  (\bibinfo{year}{2000}).

\bibitem[{\citenamefont{Buscher and Emig}(2004)}]{Buscher2004}
\bibinfo{author}{\bibfnamefont{R.}~\bibnamefont{B\"{u}scher}} \bibnamefont{and}
  \bibinfo{author}{\bibfnamefont{T.}~\bibnamefont{Emig}},
  \bibinfo{journal}{Phys. Rev. A} \textbf{\bibinfo{volume}{69}},
  \bibinfo{pages}{062101} (\bibinfo{year}{2004}).


\bibitem[{\citenamefont{Chen et~al.}(2002)\citenamefont{Chen, Mohideen,
  Klimchitskaya, and Mostepanenko}}]{Chen2002}
\bibinfo{author}{\bibfnamefont{F.}~\bibnamefont{Chen}}{\it et al.},
  \bibinfo{journal}{Phys. Rev. Lett.}
  \textbf{\bibinfo{volume}{88}}, \bibinfo{pages}{101801}
  (\bibinfo{year}{2002}).

\bibitem[{\citenamefont{Palik}(1985)}]{Palik1985}
\bibinfo  \emph{\bibinfo{title}{{\it Handbook of Optical Constants of Solids}}},
	\bibinfo{editor}{\bibfnamefont{edited by} \bibfnamefont{E.~D.} \bibnamefont{Palik}}
  (\bibinfo{publisher}{Academic Press, New York}, \bibinfo{year}{1985}).

\end{thebibliography}

\end{document}